\documentclass[final]{IEEEtran}

\usepackage{cite}
\usepackage{multirow}
\usepackage{graphicx}
\usepackage{epstopdf}
\usepackage[tight,footnotesize]{subfigure}

\usepackage{amsmath}
\usepackage{amssymb}
\usepackage{psfrag}
\usepackage{todonotes}
\renewcommand{\vec}[1]{\ensuremath{\boldsymbol{#1}}}

\begin{document}

\title{Massive MIMO performance evaluation based on measured propagation data}
\author{\IEEEauthorblockN{Xiang Gao, Ove Edfors, Fredrik Rusek, Fredrik Tufvesson}\\
\IEEEauthorblockA{Department of Electrical and Information Technology\\Lund University, Box 118, SE-22100, Lund, Sweden\\
Email: \{xiang.gao, ove.edfors, fredrik.rusek, fredrik.tufvesson\}@eit.lth.se}}

\maketitle

\begin{abstract}
Massive MIMO, also known as very-large MIMO or large-scale antenna systems, is a new technique that potentially
can offer large network capacities in multi-user scenarios.
With a massive MIMO system, we consider the case where a base station equipped with a large number of antenna elements simultaneously serves multiple single-antenna users in the same time-frequency resource.
So far, investigations are mostly based on theoretical channels with independent and identically distributed (i.i.d.) complex Gaussian coefficients,
i.e., i.i.d.~Rayleigh channels.
Here, we investigate how massive MIMO performs in channels measured in real propagation environments.
Channel measurements were performed at 2.6~GHz using a virtual uniform linear array (ULA) which has a physically large aperture, 
and a practical uniform cylindrical array (UCA) which is more compact in size, 
both having 128 antenna ports.
Based on measurement data, we illustrate channel behavior of massive MIMO in three representative propagation conditions,
and evaluate the corresponding performance.
The investigation shows that the measured channels, for both array types,
allow us to achieve performance close to that in i.i.d.~Rayleigh channels.
It is concluded that in real propagation environments we have characteristics 
that can allow for efficient use of massive MIMO, i.e.,
the theoretical advantages of this new technology can also be harvested in real channels.
\end{abstract}

\section{Introduction}\label{sec:intro}
Massive MIMO is an emerging technology in wireless communications,
which has attracted a lot of interest in recent years.
With massive MIMO, we consider multi-user MIMO (MU-MIMO) systems \cite{MU_MIMO} where base stations are equipped with a large number
(say, tens to hundreds) of antennas.
As a comparison, the LTE standard only allows for up to 8 antennas at the base station \cite{3GPP_LTE_A}.
In this way, massive MIMO scales conventional MIMO by an order or two in magnitude.
Typically, a base station with a large number of antennas serves several
single-antenna users in the same time-frequency resource. 

It has been shown in theory that such systems have potential to remarkably
improve performance in terms of link reliability, spectral efficiency, and transmit energy efficiency
\cite{Marzetta_unlimited_bs_ant, scale_up_mimo, Massive_MIMO_Larsson2013, EE_Ngo}.
Massive MIMO can also reduce intra-cell interference 
between users served in the same time-frequency resource, due to its focus of transmitted power to desired users.
The fundamental idea is that as the number of base station antennas grows large,
channel vectors between users and base station become very long random vectors 
and, under ``favorable'' propagation conditions, 
these channel vectors become pairwise orthogonal.
The term ``favorable'' is first defined in \cite{EE_Ngo} as the mutual orthogonality
among user channels,
and ``favorable'' propagation is further investigated in theory in \cite{Ngo2014_Favorable}. 
We can also interpret ``favorable'' propagation as a sufficiently complex scattering environment.
Under these conditions, even simple linear precoding/detection schemes, e.g.,
zero-forcing and matched-filtering, become nearly optimal
\cite{Marzetta_unlimited_bs_ant, scale_up_mimo, xiang_vtc}.

The attractive features of massive MIMO are, however, based on optimistic assumptions about
propagation conditions in combination with available low-cost hardware making it possible to deploy large number of antennas.
So far, investigations are mostly based on theoretical independent and identically
distributed (i.i.d.) complex Gaussian, i.e., Rayleigh fading, channels and for antenna numbers that grow without limit. 
Bringing this new technology from theory to practice, we must 
ask to what degree the optimistic theoretical predictions can be maintained in real propagation environments
when using practical antenna array setups.
In attempts to answer this question, massive MIMO propagation measurements have been conducted and measurement data used to assess massive MIMO performance in real channels \cite{xiang_vtc, eucap_linear_array, xiang_cost_2012, Xiang_Asilomar, Hoydis_meas_2012}.
Channel measurements in \cite{xiang_vtc}, at 2.6~GHz with an indoor base station using a 128-port uniform cylindrical array (UCA) of patch antennas,
showed that orthogonality of user channels improves significantly with increasing number of base station antennas. 
Already at 20 antennas, linear precoding schemes operating on measured channels achieve near-optimal performance for two users.
From measurements using a 128-element virtual uniform linear array (ULA) at 2.6~GHz, presented in \cite{eucap_linear_array} and \cite{xiang_cost_2012}, it was concluded that the angular power spectrum (APS) of the incoming waves varies significantly 
along the physically large ULA.
This is a clear indication that large-scale/shadow fading across the array is an important mechanism when dealing with physically large arrays.
As a comparison, the UCA studied in \cite{xiang_vtc} is relatively compact and much smaller in size,
but still a similar effect of variation in channel attenuations can be experienced over the array.
In this case it is not primarily a large-scale/shadow fading effect, but rather a consequence of the circular array structure and directive patch antenna elements pointing in different directions.
No matter the source of these power variations over the array, they can have a large influence on massive MIMO performance \cite{Ant_Sel_Globecom}.
A measurement campaign independent of our investigations, with an antenna array consisting of up to 112 elements, is reported in \cite{Hoydis_meas_2012}. Results obtained there, which to a large extent agree with our own experience \cite{Xiang_Asilomar},
show that despite fundamental differences between measured and i.i.d.~channels in terms of propagation characteristics, 
a large fraction of the theoretical performance gains of massive MIMO can be achieved in practice.
A different approach to characterize massive MIMO performance has been presented in \cite{VLA_Chamber_2014}, 
where real propagation environment is replaced by simulation in a reverberation chamber.

In this paper, we aim for a deeper insight into how massive MIMO performs in real propagation environments.
The investigations are based on outdoor-to-outdoor channel measurements 
using a 128-port UCA and a 128-port virtual ULA, as described in \cite{Xiang_Asilomar}.
We study the channel behavior of massive MIMO under three representative propagation conditions,
where users are:
1) closely located with line-of-sight (LOS) to the base station,
2) closely located with non-line-of-sight (NLOS) to the base station,
and 3) located far from each other.
When users are located close to each other, 
spatial multiplexing with good isolation between users can be particularly difficult,
as compared to the case when users are located far from each other. 
LOS conditions may prove particularly difficult with highly correlated channels to different users,
making spatial multiplexing less efficient. The more complex propagation in NLOS conditions is expected to decorrelate channels to different users to a larger extent.
We investigate the corresponding performance obtained in these scenarios, 
by calculating sum-rates based on measured channel data and comparing with those obtained in i.i.d.~Rayleigh channels.
As a complementary tool, we also study the singular value spreads for the measured channels.
This gives an indication of how large the difference is between the most favorable and least favorable channels. 
Small singular value spreads indicate stable channels to all users, 
while large spreads indicate that
one or more users may suffer from significantly worse conditions than others.

In this investigation we compare two different large array structures. 
From a practical point of view 
it is preferable to have a compact array, such as the UCA, since it is easier to deploy. 
However, if we make the array small, 
it will bring drawbacks such as higher antenna correlation and poor angular resolution. A two-dimensional structure like the UCA will, however, have the ability to resolve incoming waves in two dimensions.
Using a much larger one-dimensional ULA with the same number of elements, we benefit from a higher angular resolution, but only in one dimension.
Since both array structures have different characteristics, we can expect that they perform differently in a massive MIMO setting. Depending on how well the propagation environment suits each array type, one may be better than the other.
To investigate this, we compare massive MIMO performance with the
two arrays in the same propagation environments.

The rest of the paper is organized as follows. In Sec.~\ref{sec:measurements}, 
we describe our massive MIMO channel measurements and
Sec.~\ref{sec:system_description}, 
outlines the system model and performance metrics used when evaluating the measured channels, including singular value spread and sum-rate capacity.
Propagation characteristics in three measured scenarios are illustrated and discussed in Sec.~\ref{sec:propagation}.
In Sec.~\ref{sec:performance_evaluation} 
we evaluate singular value spreads and sum-rate capacities for the measured channels.
Finally, in Sec.~\ref{sec:conclusion}, we summarize our contributions and draw conclusions.

\section{Channel measurements}\label{sec:measurements}
In this section, we present the measurement campaigns for massive MIMO channels, 
on which we base our study of propagation characteristics and evaluations of system performance.
First we introduce the measurement setups, including antenna arrays and measurement equipment.
Then we describe the semi-urban environment where measurements were performed under different propagation conditions.

\subsection{Measurement setups}
\begin{figure}
  \centering
  \includegraphics[width=0.48\textwidth]{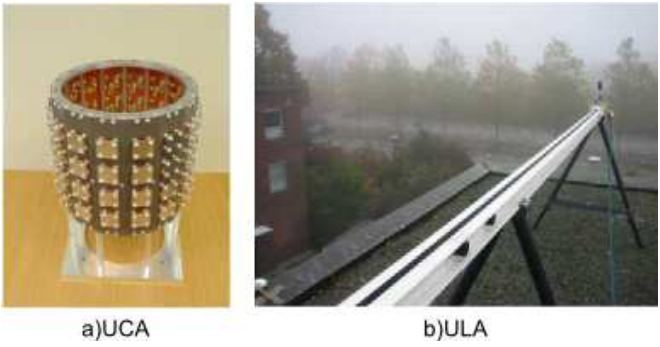}
  \caption{Two large arrays at the base station side: 
  a) a UCA with 64 dual-polarized patch antenna elements, giving 128 ports in total,
  and b) a virtual ULA with 128 vertically-polarized omni-directional antennas.}
  \label{fig:bs_array}
\end{figure}

Two channel measurement campaigns were performed with
two different large arrays at the base station side. 
Both arrays are for the 2.6~GHz range and contain 128 antenna ports each,
with antenna elements spaced half a wavelength apart. Fig.~\ref{fig:bs_array}a
shows the UCA having 64 dual-polarized patch antennas, with 16 antennas in each of the four stacked circles, 
giving a total of 128 antenna ports. 
This array is compact in size with both diameter and height around 30~cm. 
Fig.~\ref{fig:bs_array}b shows the virtual ULA with a vertically-polarized omni-directional antenna moving along
a rail, in 128 equidistant positions. 
In comparison, the ULA spans 7.4~m in space, 
which is more than 20 times the size of the UCA.
In both measurement campaigns, an omni-directional antenna with vertical polarization was used at the user side.

Channel data were recorded at center frequency 2.6~GHz and 50~MHz bandwidth.
With the UCA, measurements were taken with the RUSK LUND channel sounder, 
while for the virtual ULA, an HP~8720C vector network analyzer (VNA) was used.
With the virtual ULA and VNA,
it takes about half an hour to record one measurement, when the antenna moves from the beginning of the array to the end. 
In order to keep the channel as static as possible during one measurement, 
we performed this campaign during the night when there were 
very few objects, such as people and cars,
moving in the measurement area.
To verify that channel conditions were static enough, 
some measurements were repeated directly after the full array length was measured.
The two measurements done half an hour after each other were compared and found to match well%
\footnote{Comparing the two measured channels, i.e., the original one and the verification one,
we found that the two transfer functions are very similar, however,
there are minor differences due to channel variation and noise.
Average amplitude correlation coefficients between the two measured transfer functions over all antenna positions 
are in the range of 0.95-0.99.
Besides, we observed that the two measured channels give very similar angular power spectrum.}.

Mutual coupling among antenna elements should also be mentioned, since it is a critical issue that may affect massive MIMO performance, 
if a large number of antennas are tightly placed \cite{Taluja2013, Artiga2012}.
Although the UCA is compact, 
the worst case of mutual coupling between the neighboring elements is -11~dB \cite{IcSIK2004}.
The virtual ULA, however, experiences no mutual coupling effect.
This may lead to different performance of the virtual ULA,
as compared to a practical ULA.
However, a theoretical study in \cite{Liu_MC_2010} shows that coupling has a major impact on MIMO capacity 
only when the element separation is below 0.2 wavelengths.
Indeed, practical studies are also needed on the impact of coupling on massive MIMO performance.
This is closely related to antenna array design, a topic not covered in this paper.
We focus on the propagation aspects and investigate how different propagation conditions affect massive MIMO performance. 

\subsection{Measurement environments}
\begin{figure}
  \centering
  \includegraphics[width=0.48\textwidth]{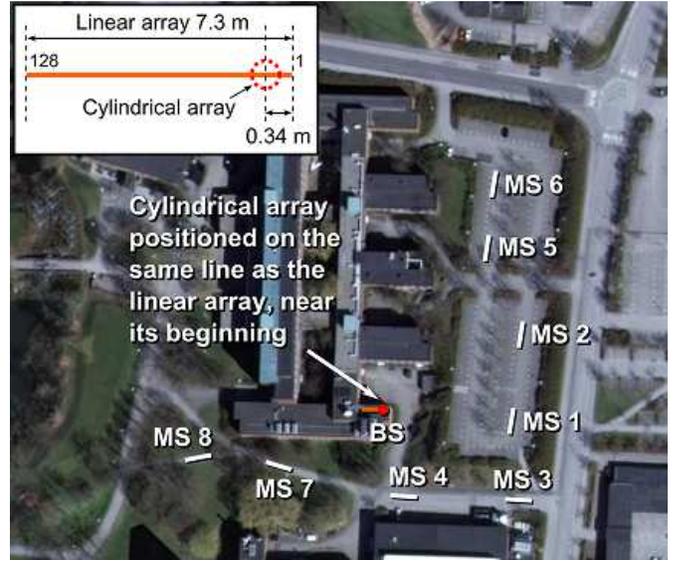}
  \caption{Overview of the measurement area at the campus of the Faculty of Engineering (LTH), Lund University, Sweden. 
At the base station side, the two antenna arrays were placed on the same roof of the E-building during their respective measurement campaigns. 
At the user side, the omni-directional antenna was moved around at MS~1-8 acting as single-antenna users.}
  \label{fig:MAP_small}
\end{figure}

The channel measurements were carried out outdoors at the E-building of
the Faculty of Engineering (LTH), Lund University, Sweden 
(N $55^{\circ}42^{\prime}37.96^{\prime\prime}$, E $13^{\circ}12^{\prime}39.72^{\prime\prime}$).
Fig.~\ref{fig:MAP_small} shows an overview of the semi-urban measurement area.
The two base station antenna arrays were placed on the same roof of
the E-building during their respective measurement campaigns.
More precisely, the position of the UCA was on the same line as the ULA, near its beginning,
and for practical reasons about 25~cm higher than the ULA.

At the user side, the omni-directional antenna was moved around the E-building at 8 measurement sites (MS) acting as single-antenna users. 
Among these sites, three (MS~1-3) have
LOS conditions, and four (MS~5-8) have NLOS conditions, while one (MS~4) has LOS
for the UCA, but the LOS component is blocked
by the roof edge for the ULA, due to the slightly lower mounting. 
Despite this, MS~4 still shows LOS characteristic for the ULA,
where one or two dominating multipath components due to diffraction at the roof edge
cause a relatively high Ricean K-factor \cite{Greenwood1999, Molisch2005}.
At MS~4, besides the roof-edge diffraction to the ULA, there is also strong scattering from the building in the south.
At each measurement site, 40 positions with about 0.5~m inter-spacing were measured with the UCA, 
and 5 positions with 0.5-2~m inter-spacing were measured with the ULA.
The reason for having fewer positions with the ULA 
was due to the long measurement time.

For all the measurements with the ULA, 
the average signal-to-noise ratio (SNR) over all antenna elements was above 28~dB,
while the lowest per-antenna SNR was above 23~dB.
With the UCA, at MS~1-4 and MS~7, the average SNR over all antenna elements was above 33~dB,
while the lowest per-antenna SNR was above 20~dB.
At MS~5-6 and MS~8, the measurement SNR was lower but still good enough,
i.e., for all antenna elements of the UCA, the SNR was about 10-25~dB.
In the measured 50~MHz bandwidth, 
we observe a coherence bandwidth about 25~MHz in the LOS scenarios, and about 5~MHz in the NLOS scenarios.

\section{System description}\label{sec:system_description}
The acquired measurement data allows study of various aspects
of massive MIMO systems.
Before discussing channel behavior and evaluating performance of massive MIMO, 
we first define our system model.

\subsection{Signal model}\label{subsec:signal_model}
We consider a single-cell multi-user MIMO-OFDM system with $N$ subcarriers in the downlink.
The base station is equipped with $M$ antennas and simultaneously serves $K$ ($K\!\leq\!M$) single-antenna users in the same time-frequency resource.
We assume that the base station has perfect channel state information (CSI),
and that the channel can be described as narrow-band at each OFDM subcarrier.

\begin{figure}
  \centering
  \includegraphics[width=0.3\textwidth]{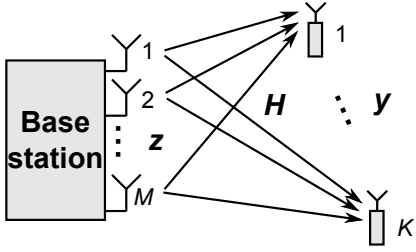}
  \caption{System model of the downlink of an MU-MIMO system with an $M$-antenna base
station and $K$ single-antenna users.}
  \label{fig:system_model}
\end{figure}

As shown in Fig.~\ref{fig:system_model}, the signal model of the considered narrow-band MU-MIMO downlink channel is
\begin{equation}\label{eq:signal_model}
	\vec{y}_\ell=\sqrt{\frac{\rho K}{M}}\vec{H}_\ell\vec{z}_\ell+\vec{n}_\ell,
\end{equation}
where $\vec{H}_\ell$ is a $K\!\times\!M$ channel matrix at subcarrier $\ell$,
$\vec{z}_\ell$ the normalized transmit vector across $M$ base station antennas, satisfying $\mathbb{E}\left\{\|\vec{z}_\ell\|^2\right\}\!=\!1$, 
$\vec{y}_\ell$ the vector of received signals at the $K$ users,
and $\vec{n}_\ell$ a complex Gaussian noise vector with i.i.d.~unit variance elements.
The term $\rho K/M$ scales the transmit energy 
and $\rho$ relates to the average per-user receive SNR%
\footnote{With the defined signal model and channel normalization, the average receive SNR at the users is smaller or equal to $\rho$,
and different values can be obtained depending on used precoding scheme. 
For example, when user channels are not completely orthogonal and inter-user interference exists,
the average receive SNR using dirty-paper coding would be higher than for zero-forcing precoding.
Equality between average per-user receive SNR and $\rho$, both for DPC and ZF precoding is obtained 
when user channels are orthogonal,
i.e., when the Gram matrix $\vec{H}_\ell\vec{H}_\ell^H$ is diagonal.}.
From the term $\rho K/M$, we increase the transmit power with
the number of users and reduce it as the number of base station antennas grows.
As $K$ increases, we keep the same transmit power per user. With increasing $M$
the array gain increases and we choose to harvest this as reduced
transmit power instead of increased receive SNR at the users%
\footnote{With realistic low-cost terminals it can be expected that only a limited SNR can be handled by the terminals, before quantization noise and dynamic range start to limit performance. 
Further, sum-rate capacities in i.i.d.~Rayleigh channels are closer to those in interference-free channels at lower SNRs\cite{scale_up_mimo}.
For these reasons, we keep a constant interference-free SNR $\rho$ at the users when the number of antennas $M$ at the
base station changes. This is to make fair and realistic comparisons of different settings.}.

Let us now return to the channel matrix $\vec{H}_\ell$ in (\ref{eq:signal_model}) and how it is formed. 
From our measurements, we have channel data obtained with 128 antenna ports at the base station
and, at the user side, each measured position represents one single-antenna user.
With the selection of $K$ positions,
we have a measured channel matrix of size $K\!\times\!128$,
which we denote $\vec{H}_\ell^\mathrm{raw}$, at subcarrier $\ell$.
The channel matrix $\vec{H}_\ell$ is then formed 
by selecting $M$ columns from a normalized version of $\vec{H}_\ell^\mathrm{raw}$.
Two different normalizations of $\vec{H}_\ell^\mathrm{raw}$ are used in different investigations. 
The two channel normalizations are:
\begin{itemize}
	\item \bfseries Normalization 1. 
	\normalfont The measured channel vectors of each user, i.e., the rows of $\vec{H}_\ell^\mathrm{raw}$, 
	denoted as $\vec{h}_{i,\ell}^\mathrm{raw},i\!=\!1,2,\ldots,K$, are normalized 
  such that the average energy over all 128 antenna ports and all $N$ subcarriers is equal to one.
  This is achieved through
	\begin{equation}\label{eq:norm_1}
		\vec{h}_{i,\ell}^\mathrm{norm} = \sqrt{\frac{128 N}{\sum\limits_{\ell=1}^N{\|\vec{h}_{i,\ell}^\mathrm{raw}\|^2}}}\vec{h}_{i,\ell}^\mathrm{raw},
	\end{equation}
	where the vector $\vec{h}_{i,\ell}^\mathrm{norm}$ is the $i$th row of the normalized channel matrix $\vec{H}_\ell^\mathrm{norm}$.
	With this normalization, the imbalance of channel attenuations between users is removed,
	while variations over antenna elements and frequencies remain.
	\item \bfseries Normalization 2. 
	\normalfont The measured channel matrix is normalized 
  such that the channel coefficients have unit average energy over all 128 antenna ports, $K$ users and $N$ subcarriers. 
  This is achieved through
	\begin{equation}\label{eq:norm_2}
		\vec{H}_\ell^\mathrm{norm} = \sqrt{\frac{128 K N}{\sum\limits_{\ell=1}^N{\|\vec{H}_\ell^\mathrm{raw}\|_\mathrm{F}^2}}}\vec{H}_\ell^\mathrm{raw},
	\end{equation}
	where
	$\|\cdot\|_\mathrm{F}$ represents the Frobenius-norm of a matrix.
	Compared with Normalization 1, here we keep the difference in channel attenuation between users,
	as well as variations over antenna elements and frequencies.
\end{itemize}
Both normalizations are done for the originally measured channel matrix with 128 columns,
rather than the matrix with $M$ columns, obtained by selecting a subset of the 128 antennas. 
The reason for this is that we would like to maintain the imbalance of channel attenuations over the antenna arrays 
due to power variations over the antenna elements. 
These variations, caused by large-scale fading/shadowing and/or directive antennas with different orientation, 
are critical for performance evaluation of massive MIMO.
When investigating singular value spreads of measured channels, we use Normalization 1.
For capacity evaluation, Normalization 2 is used in scenarios where users are closely located,
while Normalization 1 is used when users are far from each other and have large channel attenuation imbalance.
The detailed reasons for using each normalization are given in the following.

\subsection{Singular value spread}\label{subsec:svd}
As mentioned in Sec.~\ref{sec:intro},
by using a large number of antennas at the base station,
massive MIMO has the potential to separate users
so that all spatial modes are useful in such a system.
However, this relies on ``favorable'' propagation where 
user channels become pairwise orthogonal with growing number of antennas,
i.e., the off-diagonal terms of the Gram matrix $\vec{H}_\ell\vec{H}_\ell^H$ become increasingly small compared to the diagonal terms.
As this phenomenon can be easily seen in i.i.d.~Rayleigh channels, many theoretical studies are based on this assumption.
We need to investigate to what degree real massive MIMO channels are ``favorable''.
One way to evaluate joint orthogonality of all users is singular value spread of the normalized propagation matrix \cite{Gesbert2003}.
Here Normalization 1 applies, since the imbalance of channel attenuations between the users should be removed,
so that the singular value spread does not contain the difference in channel norms, but only reflects the joint orthogonality of the users.

The propagation matrix $\vec{H}_\ell$ at subcarrier $\ell$ has a singular value decomposition (SVD) \cite{Paulraj2003}
\begin{equation}\label{eq:svd}
	\vec{H}_\ell=\vec{U}_\ell\vec{\Sigma}_\ell\vec{V}_\ell^{H},
\end{equation}
where $\vec{U}_\ell$ and $\vec{V}_\ell$ are unitary matrices, 
and the $K\!\times\!M$ diagonal matrix $\vec{\Sigma}_\ell$
contains the singular values $\sigma_{1,\ell},\sigma_{2,\ell},...,\sigma_{K,\ell}$.
The singular value spread is defined as
\begin{equation}\label{eq:svd_sp}
	\kappa_\ell = \frac{\max\limits_{i}\sigma_{i,\ell}}{\min\limits_{i}\sigma_{i,\ell}},
\end{equation}
i.e., the ratio of the largest and smallest singular values.
A large $\kappa_\ell$ indicates that at least two rows of $\vec{H}_\ell$, 
i.e., the channel vectors of two users,
are close to parallel and thus relatively difficult to separate spatially,
while $\kappa_\ell\!=\!1$, i.e., 0~dB, implies the best situation where all rows are pairwise orthogonal.
The singular value spread can be an indicator whether the users should be served in the same time-frequency resource.
It also has close connection with the performance of
MIMO precoders/detectors \cite{Artes2003, Maurer2007, Mohammed2011}.

With massive MIMO, as the number of antennas increases and becomes much larger than the number of users ($M\!\gg\!K$),
we expect better orthogonality between user channels and thus smaller singular value spreads, as compared to conventional MIMO.
More importantly, we expect the singular value spread to become more stable over channel realizations.
The stability of singular value spread implies that bad channel conditions can be avoided and also leads to stability of MIMO precoders/detectors.
While the above is true for i.i.d.~Rayleigh channels, 
we investigate the measured channels in Sec.~\ref{sec:performance_evaluation},
in an attempt to find out if realistic channels also can provide sufficiently good and stable user orthogonality.

\subsection{Dirty-paper coding capacity}\label{subsec:dpc}
Through the singular value spread, 
we can investigate the potential of massive MIMO to spatially separate the users.
However, singular value spread cannot fully quantify the performance of an MU-MIMO system,
since it only offers an indication of the minimum quality of service that can be guaranteed for all users.
We would also like to know the overall performance of a massive MIMO system 
in terms of sum-rate capacity. 
A small singular value spread leads to high capacity, as interference between all users is low and
they can get relatively good quality of service.
A large singular value spread, however, does not imply a low channel capacity.
In this case, at least one user has relatively poor quality of service, 
but we do not know how many that still can get good quality of service.
For example, in a rank-deficient channel with one singular value being zero,
i.e., two user channels are aligned,
the singular value spread goes to infinity,
but the channel capacity can still be relatively high, depending on the remaining singular values.
By combining the two metrics, singular value spread and sum-rate capacity, we can get a good understanding of massive MIMO performance.

Sum-rate capacity in the narrow-band MU-MIMO downlink channel is \cite{dpc_Vishwanath2003},
\begin{equation}\label{eq:dpc_capacity}
	C_{\mathrm{DPC},\ell}=\max\limits_{\vec{P}_\ell}\log_2\det\left(\vec{I}+\frac{\rho K}{M}\vec{H}_\ell^H\vec{P}_\ell\vec{H}_\ell\right),
\end{equation}
which is achieved by dirty-paper coding (DPC) \cite{Costa1983_DPC}. 
The diagonal matrix $\vec{P}_\ell$ with $P_{\ell,i}, i\!=\!1,2,..., K$ on its diagonal allocates the transmit power among the user channels and
capacity is found by optimizing over $\vec{P}_\ell$ under the total power constraint $\sum_{i=1}^K{P_{\ell,i}}\!=\!1$. 
This can be done using the sum-power iterative water-filling algorithm presented in \cite{dpc_wf_Jindal2005}.

In measured channels where users are far from each other, 
large variations in channel attenuations to different users can have a strong influence on sum-rate capacity. 
In order to maximize the downlink sum-rate, 
a large proportion of the transmit power will be allocated to users with low channel attenuation.
These users will have relatively high date rates, compared to users with higher channel attenuation.
We can imagine an extreme case where only one user has a very high data rate and the multi-user transmission is reduced to single-user transmission.
When this happens, 
it is difficult to investigate the effect of user channel orthogonality on the system performance.
To avoid large imbalance of channel attenuations, users with similar attenuation should be grouped and served simultaneously,
while the user groups are, \emph{e.g}, time multiplexed. 
Due to a limited number of measurement positions, we do not have enough data to analyze this situation. 
We therefore focus on orthogonality between channels to different users and 
remove attenuation imbalance between users that are far apart, when evaluating their sum-rate capacity,
as described in Normalization 1.
When users are closely located, 
the path losses can be expected to be similar and any attenuation imbalance is mainly
due to small-scale and large-scale fading.
From our measurements, 
we observe that attenuation imbalance between co-located users is very small.
Thus, for capacity evaluation in this case, 
we apply Normalization 2 on the measured channels and keep the small attenuation imbalance among the users,
as is the case in i.i.d.~Rayleigh channels.

Ideally, in massive MIMO, as the number of base station antennas goes to infinity
in ``favorable'' propagation conditions, 
the channels to different users become interference free (IF) \cite{scale_up_mimo} with per-user receive SNRs approaching $\rho$ as given in our model (\ref{eq:signal_model}).
This leads to an asymptotic, interference free, sum-rate capacity
\begin{equation}\label{eq:if_capacity}
	C_\mathrm{IF}=K\log_2\left(1+\rho\right),
\end{equation}
to which i.i.d.~Rayleigh channels converge, as the number of antennas grows.
For the measured channels we would like to know how large a fraction of this capacity we can achieve.
This is investigated and discussed in Sec.~\ref{sec:performance_evaluation}.

\section{Propagation characteristics}\label{sec:propagation}
Before presenting numerical performance evaluation results,
we focus on propagation characteristics in the investigated scenarios, 
as briefly outlined in Sec.~\ref{sec:intro}.
While not providing quantitative measures of massive MIMO performance, this description gives an intuitive understanding of real massive MIMO propagation mechanisms,
and also helps to understand the evaluation results of singular value spreads and sum-rate capacities,
presented later in Sec.~\ref{sec:performance_evaluation}.
By understanding these propagation mechanisms observed in massive MIMO, 
we also gain insight into what needs to be considered and included in a massive MIMO channel model \cite{asilomar_model, Zheng_massiveMIMO_model}.

For a simple and clear illustration of massive MIMO propagation characteristics in each of the three scenarios, 
we start with four users ($K\!=\!4$), which is the number of users supported in LTE MU-MIMO \cite{3GPP_LTE_A}.
In two of the scenarios, the four users are located close to each other, with only 1.5-2~m inter-spacing,
representing situations where the spatial separation of user signals can be expected to be particularly difficult.
In the third scenario, the four users are located far from each other, with more than 10~m inter-spacing, 
representing situations where users are well distributed around the base station and we can expect good channel orthogonality.
Combining difference in user inter-spacing with LOS/NLOS conditions, the three investigated scenarios are:
\begin{enumerate}
	\item four users close to each other at MS~2, having LOS conditions to the base station,
	\item four users close to each other at MS~7, with NLOS conditions,
	\item four users far apart, at MS~1-4, respectively, all having channels with LOS characteristics.
\end{enumerate}

\begin{figure*}[h!t]
	\centering
	\subfigure[]{
		\includegraphics[width=0.35\textwidth]{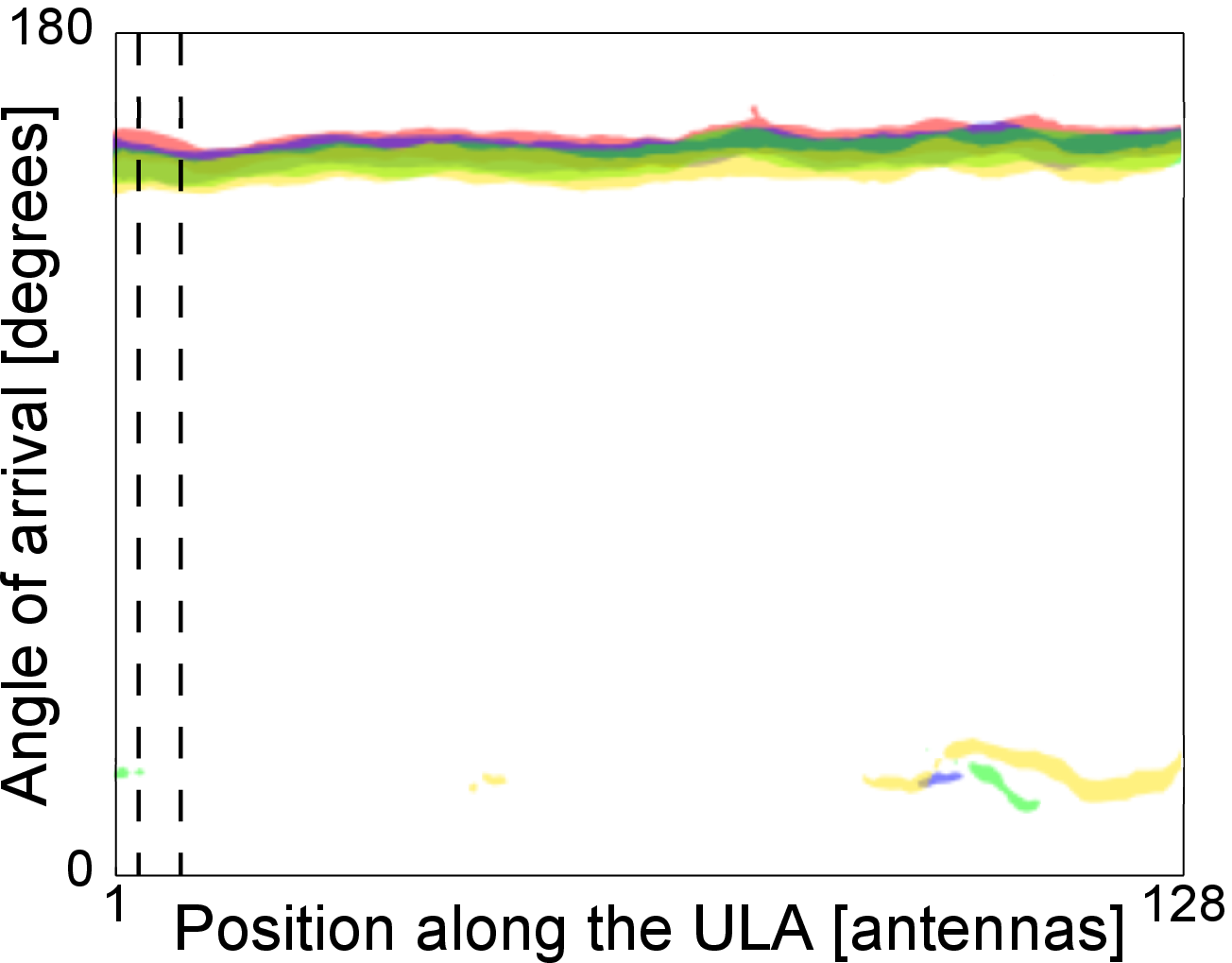}
	  \label{fig:aps_subfig1}
	 } 
	\subfigure[]{
		\includegraphics[width=0.35\textwidth]{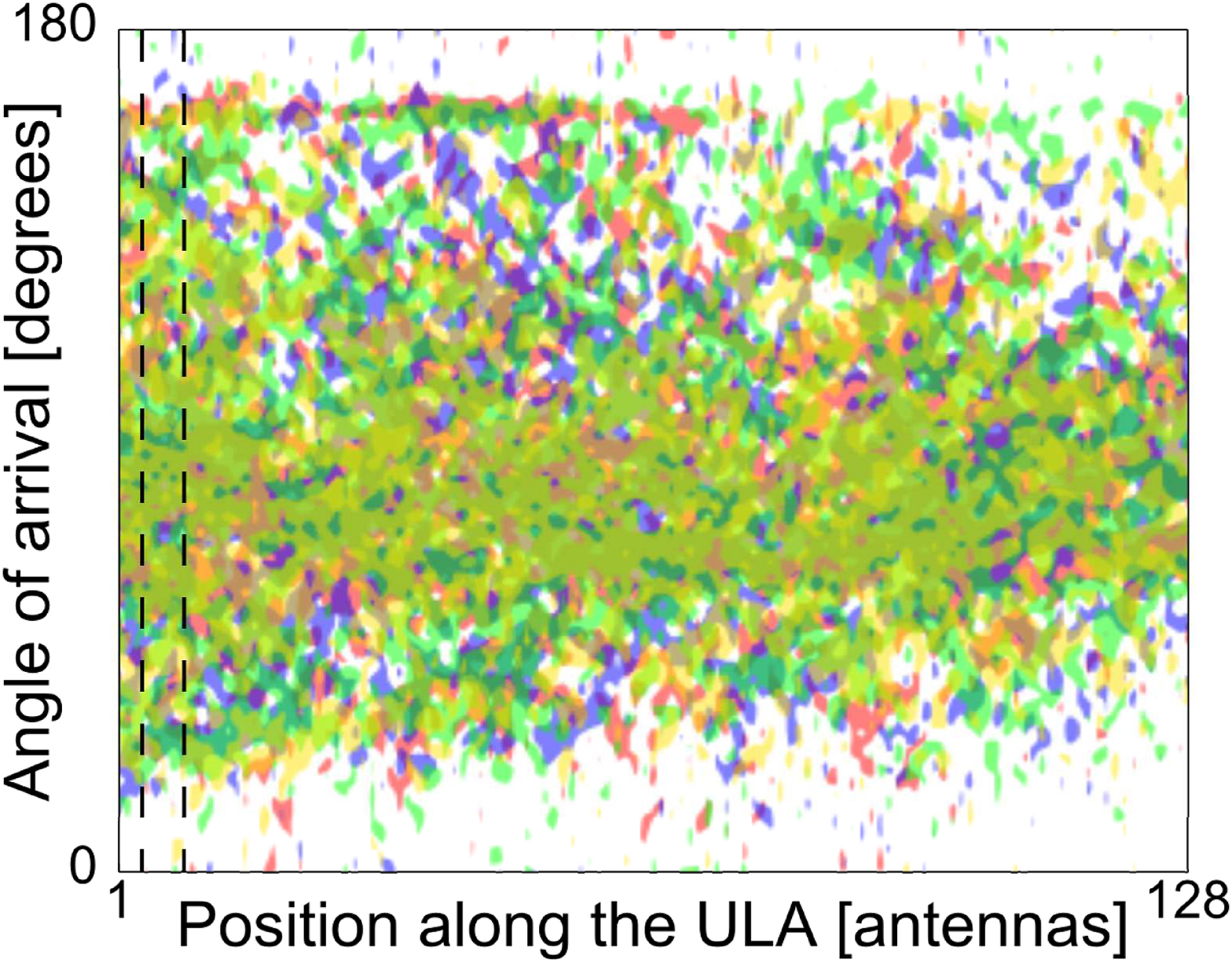}
	  \label{fig:aps_subfig2}
	 }
	\subfigure[]{
		\includegraphics[width=0.35\textwidth]{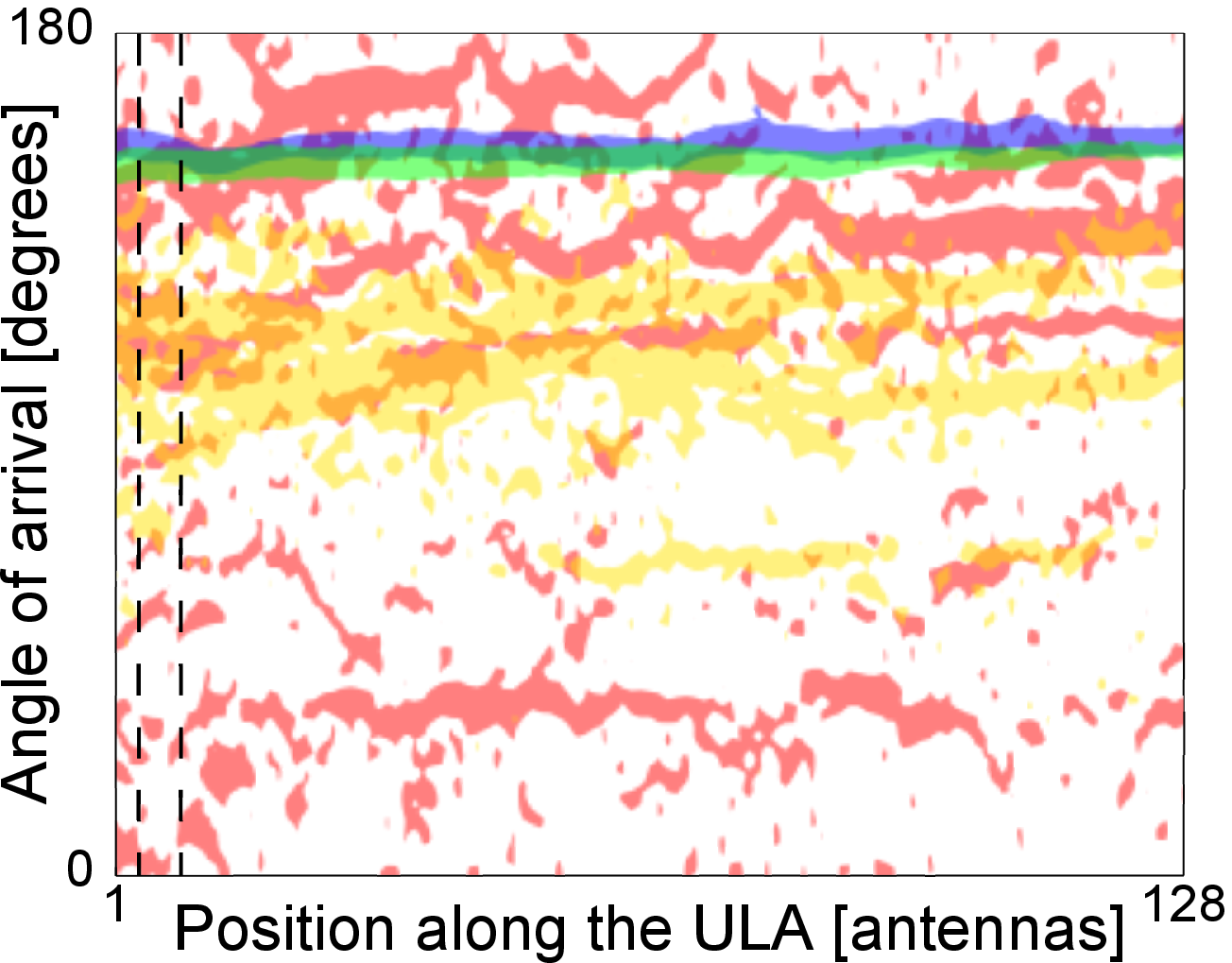}
	  \label{fig:aps_subfig3}
	 }
	\caption{Spatial fingerprints (simplified forms of the angular power spectral density along the 128-element ULA), 
in (a) a LOS scenario where the four users are co-located at MS~2,
(b) an NLOS scenario where the four users are co-located at MS~7, (c) a LOS scenario where the four users are far away from each other, at MS~1-4, respectively.
The four different colors in each plot represent the spatial fingerprints of the four different users. 
Dashed vertical lines indicate where the UCA is located and which part of the ULA propagation channel it is exposed to. 
Distinct fingerprints, as in (b) and (c), indicate relatively good conditions for spatial separation of user signals, 
while similar fingerprints, as in (a), indicate that spatial separation may be more difficult.}
	\label{fig:aps}
\end{figure*}

With the aim of assisting understanding of the physical propagation channels, 
we estimate the APS at the base station side.
The directional estimates for the ULA are obtained through the space-alternating generalized expectation maximization
(SAGE) algorithm \cite{Fleury1999_SAGE}, which jointly estimates the delay, 
incidence azimuth, and complex amplitude, of multipath components (MPCs) in radio channels.
The frequency-dependent SAGE algorithm is applied to a sliding window of 10 neighboring elements on the ULA
and, for the measured channel within each window, 200 MPCs are estimated.
The reason for estimating the MPC parameters based on 
10-antenna windows is that the incoming waves can be considered planar if the array is small enough. This aperture
corresponds to a Fraunhofer distance of about 5~m, making waves originating from reflections beyond that distance
to appear planar. 
Using 10 antennas also provides a relatively high angular resolution for the directional estimation. 
Note that the range of azimuth estimation is 0-180 degrees for the ULA,
due to inherent directional ambiguity problem when using this type of array structure \cite{Ambiguity_LinearArray}.

Based on the SAGE estimates, we obtain the APS in azimuth at each position along the ULA.
For each scenario, we compare the APS from different users as seen at the base station.
For the convenience of comparison, we simplify the APS from each user.
Instead of showing the estimated power levels from all the azimuth directions,
we only show 
from which directions the incoming energy is strongest. The colored patches show where 90\% of the total energy across the whole array is concentrated. 
This simplified form of APS illustrates the directional pattern of the incoming energy from a specific user.
Since it is a highly simplified form of the spatial properties of the channel from a specific user, we call it a ``spatial fingerprint''.
In Fig.~\ref{fig:aps}, for each scenario, we plot the four users' spatial fingerprints on top of each other.
The four colors in each plot represent the spatial fingerprints of the four different users.
Since the UCA was positioned at the beginning of the ULA, 
as indicated by the dashed lines in Fig.~\ref{fig:aps}, 
we consider that it experiences the propagation channels at that particular part of the ULA,
but with directional patch antennas oriented in different directions.

What can we learn from these spatial fingerprints?
First, they provide an intuitive understanding of the distribution of incoming energy from different users in real channels under different propagation conditions.
Secondly, by comparing fingerprints,
we get an understanding of how much the APS changes between users in different scenarios.
Through this we can acquire qualitative information about how difficult it is to do spatial separation of signals from different users.
Distinct fingerprints indicate relatively good spatial orthogonality of the user channels, 
and we can expect that the user signals can be separated with rather simple means.
In this case, the channels have relatively small singular value spreads and relatively high sum-rate capacities.
Similar and overlapping fingerprints, on the contrary,
represent a more difficult situation and spatial separation of user signals may be much harder.
With incoming energy from largely the same directions, detailed knowledge about amplitude and phase
is needed to fully assess the situation.
Thirdly, these fingerprints allow for a direct comparison of the propagation channels experienced by massive and more conventional MIMO systems.
This can be done by comparing fingerprints along the entire ULA with the local fingerprint somewhere along the ULA, that would be experienced by a smaller conventional MIMO array.
Lastly, an attempt to develop a sophisticated geometry-based channel model for massive MIMO should likely take 
these spatial fingerprints into consideration. 
Our point of view is that if a channel model does not reflect the spatial properties observed through these fingerprints, 
it does not accurately model the nature of a massive MIMO propagation channel.
We discuss these issues in the following.

First we turn our attention to propagation conditions and spatial separability of user channels in the three investigated scenarios.
In Fig.~\ref{fig:aps}a, we can see that in the LOS scenario with co-located users, 
incoming energy from all users is concentrated around 160 degrees, which is the LOS direction.
For some users, a significant amount of energy also comes from some scatterers at around 20 degrees at the end of the ULA.
The overlap of the four users' fingerprints indicates that we may have a relatively high correlation%
\footnote{The spatial correlation we talk about here is an instantaneous property between users,
rather than an average property, e.g., over time realizations of the channels.}
between their channels, 
making it difficult to spatially separate signals from the co-located users.
However, as discussed above, amplitude and phase differences may still make users easier to separate than they appear from studying the fingerprints.

An entirely different situation is shown Fig.~\ref{fig:aps}b, 
where the four users are still closely located but in an NLOS scenario with rich scattering.
Incoming energy from all four users is distributed over a much larger angle across the whole array,
reflecting a rich scattering environment.
The four users' fingerprints are very complex
and quite different from each other, as compared to the case in Fig.~\ref{fig:aps}a.
This indicates that the spatial correlation between channels to the users is relatively low, 
which should allow for a better spatial separation of user signals, even though they are still closely located.

Fig.~\ref{fig:aps}c shows the scenario where four users are located far from each other, all having LOS propagation characteristics.
Users at MS~2 and 3, whose fingerprints are in blue and green, respectively, have the strongest LOS characteristics with
incoming energy concentrated to a certain direction along the entire array.
This is in stark contrast to users at MS~1 and 4, whose fingerprints are in red and yellow.
At MS~1, the LOS is at the end-fire direction of the ULA, 
and its power contribution is weakened due to the superposition with the ground reflection.
At MS~4, besides the energy from the roof-edge diffraction to the ULA, 
strong scattering from the building in the south also contributes considerably.
Since the users are located at different sites, their fingerprints should be very different from each other.
Note that the signals from users at MS~2 and 3 appear to come from the same direction due to the inherent angular ambiguity of the ULA.
However, as seen later in Sec.~\ref{sec:performance_evaluation} it is possible to spatially separate the two users.
A good spatial separation of all users can be expected in this scenario.

Now, let us turn our attention to propagation channels experienced by massive and more conventional MIMO systems.
In the fingerprint plots, we can see that the ULA potentially experiences channels with much more spatial variations,
as compared to small arrays spanning only a few wavelengths in space.
Large spatial variations can help to decorrelate channels even when users are closely located, 
as in Fig.~\ref{fig:aps}b.
Fingerprints may overlap locally,
but over longer distances along the array they are quite distinct.
This indicates that, with small arrays users may have relatively low spatial correlation on average, e.g., over time,
while with a physically large ULA 
decorrelation of user channels can be instantaneous.
However, strong LOS may reduce the ability of the ULA to spatially separate signals from co-located users,
such as the situation shown in Fig.~\ref{fig:aps}a.
Since we do not consider the phase information over the array there,
we later investigate this situation in more detail by evaluating both singular value spreads and sum-rate capacities.

For the compact UCA, experiencing only a small part of channels seen by the ULA,
separation of user signals may be more difficult.
When users are closely located and incoming energy is concentrated to similar and narrow directions, 
patch antennas oriented in ``wrong'' directions may have high channel attenuations and contribute little to spatial separation of signals from co-located users.
Despite this, the UCA may still gain from its circular structure and provide good user decorrelation,
when users are distributed around the base station, 
and incoming energy is distributed in different directions, 
as shown in Fig.~\ref{fig:aps}c.

\section{Performance evaluation}\label{sec:performance_evaluation}
To get a more quantitative understanding of how massive MIMO would perform in our measured channels,
we turn our attention to singular value spreads and sum-rate capacities in the three measured scenarios. 
First we focus on the case of four users ($K\!=\!4$),
as we did in the propagation characteristics in Sec.~\ref{sec:propagation}.
We then increase the number of users to sixteen ($K\!=\!16$) and investigate the performance when
more users are served simultaneously.

\subsection{Four users ($K\!=\!4$)}
In all three scenarios,
over $N\!=\!161$ subcarriers and 2000 random selections of antenna subsets,
i.e., selections of $M$ antennas out of the 128,
we show a) the cumulative distribution functions (CDFs) of the singular value spreads in the channels, 
when using 4, 32 and 128 base station antennas,
and b) the average DPC capacities including their 90\% confidence intervals,
as the number of base station antennas $M$ grows from 4 to 128.
Note that for $M\!=\!128$ there is only a single choice of selecting the antenna subset, 
and the CDFs of the singular value spreads and the capacity confidence intervals
are therefore computed over frequencies only. 
For $M\!<\!128$, as the number of all possible antenna subsets can be extremely large,
we randomly select 2000 subsets, 
and let the CDFs of the singular value spreads and the capacity confidence intervals also take the random antenna selections into account.
As a reference, we also show simulated results for i.i.d.~Rayleigh channels.
We select the interference-free SNR to $\rho\!=\!10$~dB
\footnote{The performance of i.i.d.~Rayleigh channels at different SNR levels has been derived in \cite{scale_up_mimo}.
We select the interference-free SNR to be 10~dB since it is a middle-level SNR.}, 
and with four users the asymptotic capacity (\ref{eq:if_capacity}) becomes $4\log_2\left(1\!+\!10\right)\!=\!13.8$~bps/Hz.
In the following we discuss the singular value spreads and DPC capacities in the three scenarios.

\subsubsection{Four users co-located with LOS}
\begin{figure}
  \centering
  \includegraphics[width=0.48\textwidth]{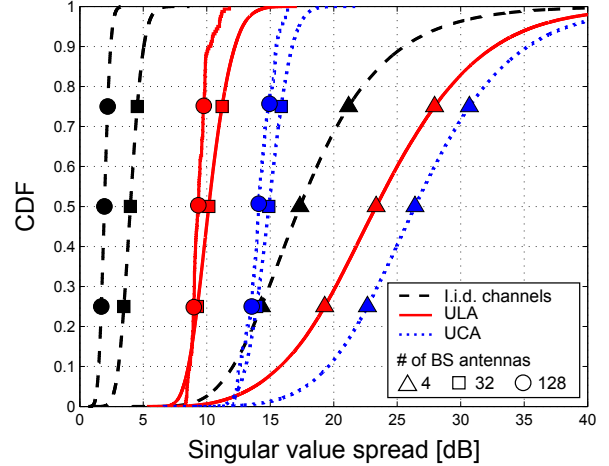}
  \caption{CDFs of singular value spreads when using 4, 32 and 128 antennas, 
in the scenario where the four users are closely located at MS~2, all having LOS to the base station antenna arrays.}
  \label{fig:svd_los_co}
\end{figure}

As discussed in Sec.~\ref{sec:propagation}, this scenario represents a particularly difficult situation 
for spatial separation of user signals,
which can be seen from the four users' similar fingerprints in Fig.~\ref{fig:aps}a.
First we study the CDFs of singular value spreads, as shown in Fig.~\ref{fig:svd_los_co}.
We observe that for i.i.d.~Rayleigh channels, the median of the singular value spread significantly reduces from 17~dB to below 4~dB, 
as the number of antennas increases from 4 to 32 and 128. 
Singular value spreads also become much more stable around small values, 
as the CDF curves have no substantial upper tails.

For the measured channels, using either ULA and UCA, 
the singular value spreads are significantly larger than those of i.i.d.~Rayleigh channels,
for all three numbers of antennas.
This indicates a much worse user channel orthogonality in the measured channels,
due to co-location of users and strong LOS conditions in this scenario.
Still, trends similar to those seen in i.i.d.~Rayleigh channels can be observed in the measured channels. 
The median of the singular value-spread decreases by 14 dB with the ULA and 12 dB with the UCA,
as the number of antennas increases from 4 to 128.
Meanwhile, when using a large number of antennas,
the substantial upper tails of the CDF curves reduce, and almost disappear in the case of 128 antennas.
With only 4 antennas, the selections of antenna subsets and subcarriers can make a big difference on the user orthogonality.
This means that with small arrays we may encounter propagation channels
with very good conditions as well as very bad ones,
depending on the choice of antenna positions and used subcarriers.
When increasing the number of antennas to 32,
user orthogonality improves and becomes much more stable over antenna selections and subcarriers.
Thus, bad channel conditions can largely be avoided by adding more antennas at the base station.
When using all 128 antennas, user orthogonality improves further and becomes more stable over subcarriers.
The above observations tell us that despite a significant gap between measured and i.i.d.~Rayleigh channels in this scenario,
spatial separation of signals from co-located users can be greatly improved by using a large number of antennas,
and more importantly, the results become more stable over both subcarriers and different antenna selections.

\begin{figure}
  \centering
  \includegraphics[width=0.48\textwidth]{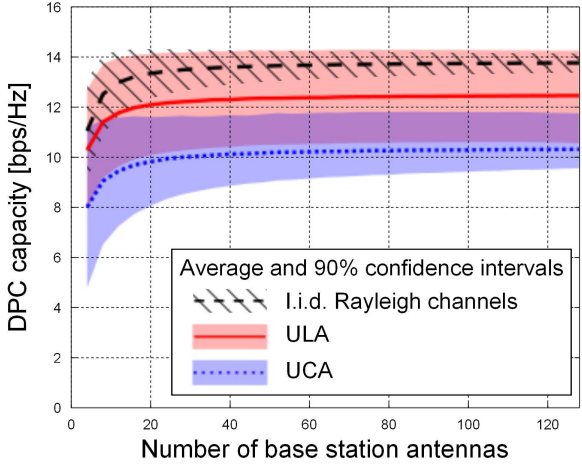}
  \caption{Sum-rate capacity in the downlink, achieved by DPC, in a scenario where four users are close to each other at MS~2, all having LOS to the base station antenna arrays.}
  \label{fig:dpc_capacity_los_co}
\end{figure}

We now move to sum-rate capacities achieved by DPC, as shown in Fig.~\ref{fig:dpc_capacity_los_co}. 
As a reference, the average capacity in i.i.d.~Rayleigh channels converges
to the asymptotic capacity value of 13.8~bps/Hz and the capacity variation
becomes smaller as the number of antennas increases.
In the measured channels, however, averages are significantly lower and variations are larger.
Let us focus on the average capacities first, and discuss the variations later.
The drops in average capacities for measured channels coincide with larger singular value spreads.
Despite this, in this potentially difficult spatial separation situation,
the ULA and UCA perform at 90\% and 75\% of the asymptotic capacity, respectively, 
when the number of antennas is above 40, i.e., when the number of antennas is 10 times the number of users.

\subsubsection{Four users co-located with NLOS}
\begin{figure}
  \centering
  \includegraphics[width=0.48\textwidth]{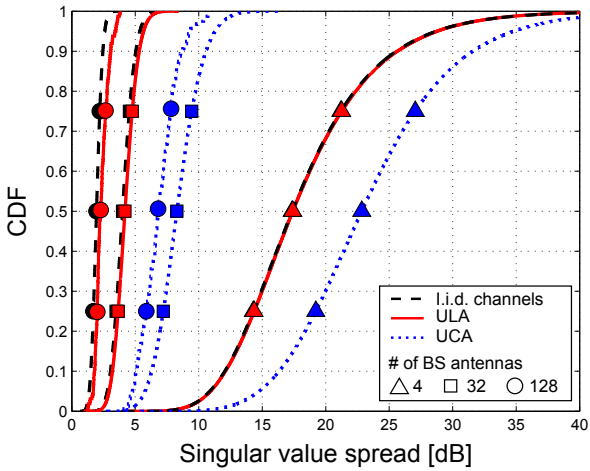}
  \caption{CDFs of singular value spreads when using 4, 32 and 128 antennas, 
in the scenario where four users are closely located at MS~7, with NLOS to the base station antenna arrays.}
  \label{fig:svd_nlos_co}
\end{figure}

In this scenario we still have users closely located, but now in NLOS conditions. 
NLOS with rich scattering, 
as illustrated in Fig.~\ref{fig:aps}b, 
where spatial fingerprints of users are complex and distinct, 
should improve the situation by providing
more ``favorable'' propagation and thus allowing better spatial separation of user signals.
The benefits of complex propagation are reflected in the CDFs of singular value spreads in Fig.~\ref{fig:svd_nlos_co}.
Singular value spreads in this scenario become significantly smaller,
as compared to those in the corresponding LOS case.
Especially for the ULA, the CDF curves are very close to those of i.i.d.~Rayleigh channels.
The substantial upper tails of the CDF curves observed when using a small number of antennas
disappear when using all 128 antennas in the measured channels.
This means that over the measured bandwidth the probability of seeing a singular value spread much larger than 
2~dB for the ULA, and 7~dB for the UCA, is very low.

\begin{figure}
  \centering
  \includegraphics[width=0.48\textwidth]{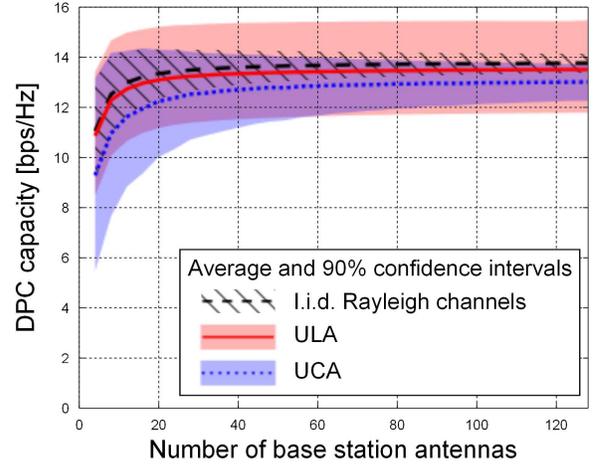}
  \caption{Sum-rate capacity in the downlink, achieved by DPC, in the scenario where the four users are close to each other at MS~7, with NLOS to the base station antenna arrays.}
  \label{fig:dpc_capacity_nlos_co}
\end{figure}

Correspondingly, the benefits brought by the NLOS condition with rich scattering can also be observed in DPC capacities, as shown in Fig.~\ref{fig:dpc_capacity_nlos_co}.
Despite co-located users, the ULA here provides average performance very close to the asymptotic capacity achieved in i.i.d.~Rayleigh channels, 
while the UCA reaches more than 90\%,
when the number of antennas is above 40.

\subsubsection{Four users located far from each other with LOS}
\begin{figure}
  \centering
  \includegraphics[width=0.48\textwidth]{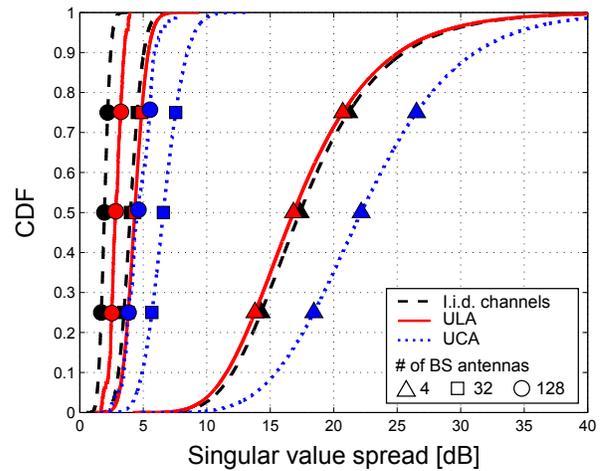}
  \caption{CDFs of singular value spreads when using 4, 32 and 128 antennas, 
in the scenario where four users are well separated at MS~1-4, respectively, with LOS characteristics.}
  \label{fig:svd_los_sp}
\end{figure}

In this scenario, despite LOS characteristics, 
increased inter-spacing between users should help to improve performance.
As can be seen in Fig.~\ref{fig:aps}c, the users' spatial fingerprints are reasonably different,
which indicates a favorable decorrelation situation between user channels for the large arrays.
In the CDFs of singular value spreads shown in Fig.~\ref{fig:svd_los_sp}, 
the ULA again performs very close to i.i.d.~Rayleigh channels.
The UCA has a significant improvement as compared to the two previous scenarios: 
the median of the singular value spread reduces to below 5 dB when using 128 antennas.
Singular value spreads in the measured channels again become quite stable when using a large number of antennas.

\begin{figure}
  \centering
  \includegraphics[width=0.48\textwidth]{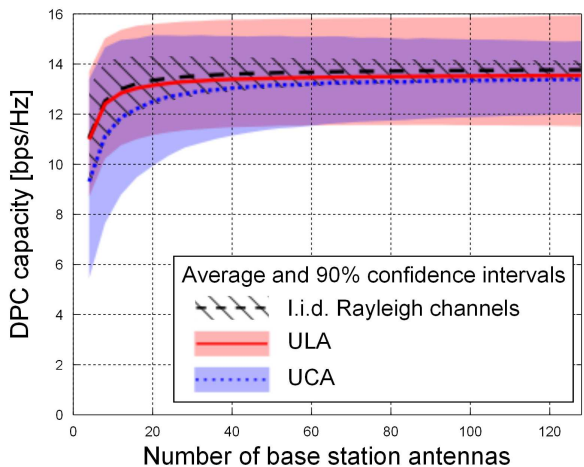}
  \caption{Sum-rate capacity in the downlink, achieved by DPC, in the scenario where the four users are well separated at MS~1-4, respectively, with LOS characteristics.}
  \label{fig:dpc_capacity_los_sp}
\end{figure}

As can be seen in Fig.~\ref{fig:dpc_capacity_los_sp},
both the ULA and the UCA perform very close to that of the asymptotic capacity achieved in i.i.d.~Rayleigh channels,
when having more than 40 antennas.
The UCA shows slightly lower performance than the ULA.

Throughout the three scenarios discussed above and whose
performances are shown in Fig.~\ref{fig:svd_los_co} - Fig.~\ref{fig:dpc_capacity_los_sp},
we observe that the ULA performs better than the UCA.
Due to its large aperture, the ULA experiences more spatial variations in the channels over the array,
which provide better distinction between user channels and thus better spatial separation.
In other words, the ULA has a very high angular resolution,
which helps it resolve scatterers better than the compact UCA.
The small aperture of the UCA and its patch antennas facing different directions 
make it difficult to resolve scatterers at similar azimuth angles, which is usually the case when users are located close to each other.
When users are well distributed around the base station,
the UCA can separate scatterers at different azimuth angles, 
and achieves better performance.

For the DPC capacities, we focused on averages in the previous discussions. 
Now we turn our attention to the variations over frequencies and random antenna selections.
Comparing with i.i.d.~Rayleigh channels, we notice that capacity variations in measured channels are much larger,
and decrease much slower as the number of antennas increases. 
This is due to larger power variations over antenna elements and over frequencies in the measured channels.
For the ULA, power variation over antenna elements is due to large-scale/shadow fading experienced across the array, 
as reported in \cite{eucap_linear_array, xiang_cost_2012},
while for the UCA, it is mainly due to its circular structure with directional patch antennas oriented differently.
With omni-directional antenna elements, the ULA has larger power variations over the measured bandwidth,
as compared to the UCA with directional antenna elements. 
This gives the ULA larger capacity variations than the UCA,
especially in the case of 128 antennas when the capacity variations are only across frequencies.
Note that although the average capacity increases with the number of antennas,
for some antenna selections a small number of antennas can perform better than a larger number of antennas.
This can be observed from the upper part of the 90\% confidence intervals of the UCA in Fig.~\ref{fig:dpc_capacity_nlos_co}.
This is because in our signal model we reduce the transmit power with increasing number of antennas,
while some antennas contribute more to the capacity than the others.
It implies that we may gain by selecting the ``right'' antennas, as discussed in \cite{Ant_Sel_Globecom}.

In all three scenarios with four users and the ULA, 
as few as 20 antennas gives very competitive performance, 
while slightly higher numbers are required for the UCA.
However, when using more practical precoding schemes,
such as zero-forcing (ZF) and matched-filtering (MF) precoding, 
sum-rate converges slower, 
which means that more antennas are needed to achieve the required performance.
This is shown in \cite{Massive_MIMO_Larsson2013} and \cite{Xiang_Asilomar}.
More antennas are also needed, if we want to serve more users in the same time-frequency resource.

\subsection{Sixteen users ($K\!=\!16$)}
While only four users are supported in LTE MU-MIMO, 
with more than one hundred antennas at the base station,
massive MIMO can potentially serve many more users simultaneously.
Here we increase the number of users to sixteen ($K\!=\!16$),
and again investigate singular value spreads and achieved sum-rate capacities. 
Due to limited number of measurement positions with the ULA, we concentrate on the UCA.
In the two scenarios with co-located users, 
we simply increase the number of users from 4 to 16, and the inter-spacing between users is about 0.5~m.
In the scenario where users are located far from each other, we select two users from each of the sites MS~1-8, 
with an inter-spacing larger than 10~m. Doing so, eight users have LOS conditions while the other eight have NLOS.

\begin{figure}
  \centering
  \includegraphics[width=0.48\textwidth]{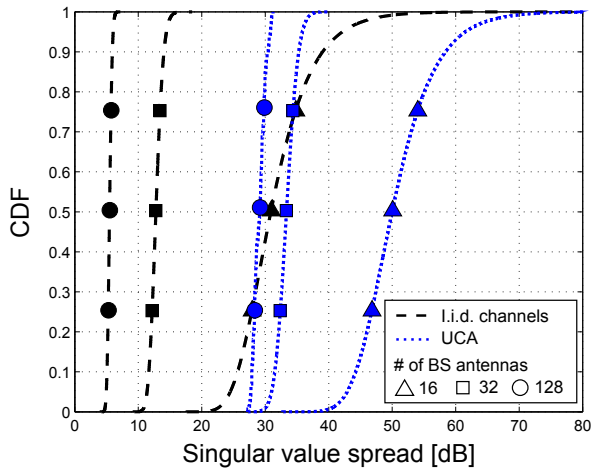}
  \caption{CDFs of singular value spreads when using 16, 32 and 128 antennas, 
in the scenario where sixteen users are closely located at MS~2, all with LOS to the UCA.}
  \label{fig:svd_los_co_16_user}
\end{figure}

\begin{figure}
  \centering
  \includegraphics[width=0.48\textwidth]{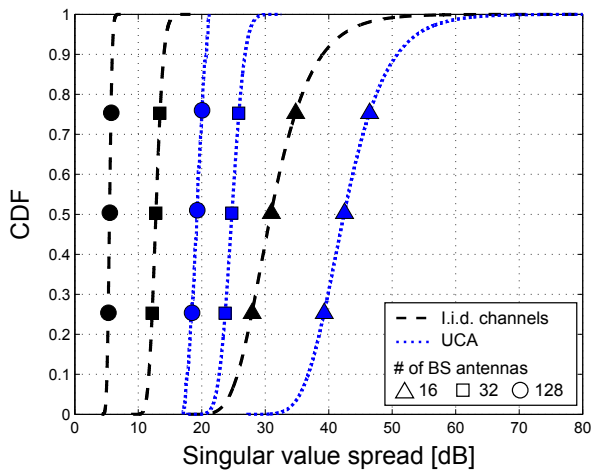}
  \caption{CDFs of singular value spreads when using 16, 32 and 128 antennas, 
in the scenario where sixteen users are located close to each other at MS~7, all with NLOS to the UCA.}
  \label{fig:svd_nlos_co_16_user}
\end{figure}

\begin{figure}
  \centering
  \includegraphics[width=0.48\textwidth]{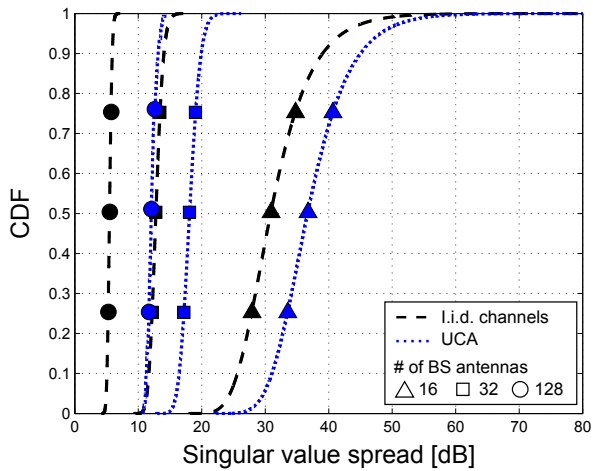}
  \caption{CDFs of singular value spreads when using 16, 32 and 128 antennas, 
in the scenario where sixteen users are located far from each other at MS~1-8, among which eight have LOS conditions and
eight have NLOS to the UCA.}
  \label{fig:svd_los_nlos_sp_16_user}
\end{figure}

CDFs of singular value spreads in the three scenarios are shown in 
Fig.~\ref{fig:svd_los_co_16_user} - Fig.~\ref{fig:svd_los_nlos_sp_16_user}.
In both measured channels and i.i.d.~Rayleigh channels, singular value spreads are larger 
than those in the four user cases.
This indicates, as expected, that with more users it is more difficult to
spatially separate their signals.
In the scenario where sixteen users are co-located with LOS, 
as shown in Fig.~\ref{fig:svd_los_co_16_user},
singular value spreads are much larger than those in i.i.d.~Rayleigh channels.
The situation improves significantly in the NLOS scenario, as shown in Fig.~\ref{fig:svd_nlos_co_16_user}. 
The gap in singular value spreads between measured and i.i.d.~Rayleigh channels becomes smaller,
which again indicates that NLOS with rich scattering provides more ``favorable'' propagation for the spatial separation of user signals,
even when they are located close to each other.
When sixteen users are located far from each other,
the CDF curves of singular value spreads in the measured channels are closer to the ones for i.i.d.~Rayleigh channels, as shown in Fig.~\ref{fig:svd_los_nlos_sp_16_user}.
This implies that spatial separation of user signals improves even more.
In all three scenarios, despite larger singular value spreads in the measured channels,
trends similar to those for i.i.d.~Rayleigh channels can be observed.
The singular value spread becomes smaller and much more stable, as the number of base station antennas increases.

\begin{figure}
  \centering
  \includegraphics[width=0.48\textwidth]{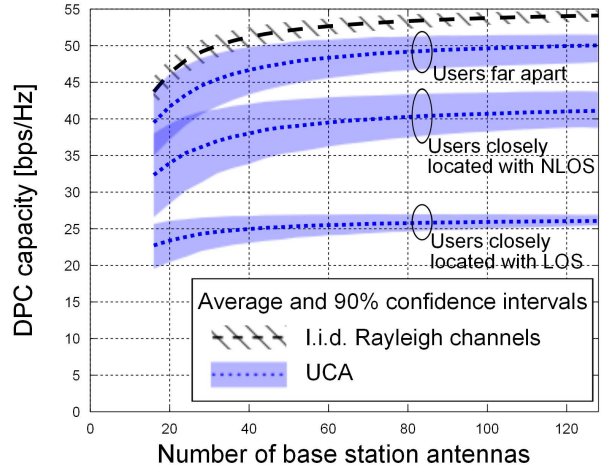}
  \caption{Sum-rate capacity in the downlink, achieved by DPC, 
in the scenario where sixteen users are located close to each other at MS~2 with LOS, MS~7 with NLOS, and are far from each other at MS 1-8, respectively.}
  \label{fig:dpc_capacity_16_users}
\end{figure}

DPC capacities in the three scenarios are shown in Fig.~\ref{fig:dpc_capacity_16_users}.
With sixteen users, asymptotic capacity given in (\ref{eq:if_capacity}) is $16\log_2\left(1\!+\!10\right)\!=\!55.4$~bps/Hz.
Average performance in i.i.d.~Rayleigh channels gets closer and closer to this asymptotic capacity,
as the number of antennas increases.
Performance in the measured channels is, however, significantly lower.
Despite this, in the worst case where sixteen users are co-located with LOS, 
the average performance reaches about 50\% of the asymptotic capacity when all 128 antennas are used, i.e., 8 times the number of users.
The situations in the other two scenarios are better. 
With 128 antennas, the UCA performs at 75\% and 90\% of the asymptotic capacity,
when sixteen users are co-located with NLOS and are far apart, respectively.

With more users and equal number of base station antennas, spatial separation becomes more difficult, 
but with the UCA we still obtain a large fraction of the i.i.d.~Rayleigh performance,
especially in NLOS conditions with rich scattering and when users are far apart.
Although we lack measurement data for sixteen users with the ULA,
we can expect that the ULA would provide better spatial separation also in this case, especially for co-located users,
due to its higher angular resolution. 

\section{Summary and conclusions}\label{sec:conclusion}
The presented investigation shows that in the studied real propagation environments 
we have characteristics that allow for efficient use of massive MIMO:
the advantages of this new technology, as predicted by theory, can also be obtained in real channels.
Based on channel measurements, using one practical UCA and one virtual ULA, both having 128 elements,
we have illustrated the channel behavior of massive MIMO in three representative propagation scenarios
and discussed corresponding singular value spreads and achieved sum-rate capacities.

In all scenarios,
the singular value spread decreases considerably,
and becomes more stable around a smaller value over the measured bandwidth, 
when using a large number of antennas.
This indicates that massive MIMO provides better orthogonality between channels to different users and 
better channel stability than conventional MIMO.
In the most difficult situation studied, i.e., closely located users with strong LOS to the base station, 
the singular value spread is significantly larger than that in i.i.d.~Rayleigh channels,
which indicates worse user orthogonality in the measured channels.
Despite this gap,
a large fraction of the asymptotic capacity achieved in i.i.d.~Rayleigh channels can still be harvested in the measured channels.
In the other studied scenarios,
NLOS conditions with rich scattering provide more ``favorable'' propagation and allow better spatial separation of the users,
even though they are closely located,
while well distributed users also help to improve the performance.
In the scenarios where users are in NLOS or in LOS but located far from each other, 
the measured channels with the ULA and the UCA achieve performance close to that in i.i.d.~Rayleigh channels.

\section*{Acknowledgement}
The authors would like to acknowledge the support from ELLIIT - an Excellence Center at Link{\"o}ping-Lund in Information Technology,
and the Swedish Research Council (VR) as well as the Swedish Foundation for Strategic Research (SSF). 
The research leading to these results has received funding from the European Union Seventh Framework Programme (FP7/2007-2013)
under grant agreement n$^{\circ}$ 619086 (MAMMOET).

\bibliographystyle{IEEEtran}
\bibliography{bib/IEEEabrv,bib/asilomar_journal_refs}

\end{document}